\documentclass[prb, 11pt,superscriptaddress,floatfix,nofootinbib,notitlepage,longbibliography]{revtex4-1}
\usepackage{graphicx}
\usepackage[colorlinks=true,citecolor=blue]{hyperref}
\usepackage{amsmath}
\usepackage{amssymb}
\usepackage{soul}
\usepackage{gensymb}
\usepackage{amsfonts,amssymb,amsbsy}
\usepackage{siunitx}
\usepackage{float}
\usepackage{physics}
\usepackage{booktabs}
\usepackage[percent]{overpic}

\newcommand{\dIdV}{d$I$/d$V$}
\usepackage{xcolor}

\usepackage{lineno}

\begin{document}

\title{Control of molecular orbital ordering using a van der Waals monolayer ferroelectric}

\author{Mohammad Amini}
\affiliation{Department of Applied Physics, Aalto University, FI-00076 Aalto, Finland}

\author{Orlando J. Silveira}
\affiliation{Department of Applied Physics, Aalto University, FI-00076 Aalto, Finland}

\author{Viliam Va\v{n}o}
\affiliation{Department of Applied Physics, Aalto University, FI-00076 Aalto, Finland}

\author{Jose L. Lado}
\affiliation{Department of Applied Physics, Aalto University, FI-00076 Aalto, Finland}

\author{Adam S. Foster}
\affiliation{Department of Applied Physics, Aalto University, FI-00076 Aalto, Finland}
\affiliation{WPI Nano Life Science Institute (WPI-NanoLSI), Kanazawa University, Kakuma-machi, Kanazawa 920-1192, Japan}

\author{Peter Liljeroth}
\affiliation{Department of Applied Physics, Aalto University, FI-00076 Aalto, Finland}

\author{Shawulienu Kezilebieke}
\email{kezilebieke.a.shawulienu@jyu.fi}
\affiliation{Department of Physics, Department of Chemistry and Nanoscience Center, 
University of Jyväskyl\"a, FI-40014 University of Jyväskyl\"a, Finland}

\date{\today}

\begin{abstract}

Two-dimensional (2D) ferroelectric materials provide a promising platform for the electrical control of quantum states. In particular, due to their 2D nature, they are suitable for influencing the quantum states of deposited molecules via the proximity effect. Here, we report electrically controllable molecular states in phthalocyanine molecules adsorbed on monolayer ferroelectric material SnTe. In particular, we demonstrate that the strain and ferroelectric order in SnTe creates a transition between two distinct orbital orders in the adsorbed phthalocyanine molecules. By controlling the polarization of the ferroelectric domain using scanning tunneling microscopy (STM), we have successfully demonstrated that orbital order can be manipulated electrically. Our results show how ferroelastic coupling in 2D systems allows control of molecular states, providing a starting point for ferroelectrically switchable molecular orbital ordering and ultimately, electrical control of molecular magnetism.


\end{abstract}

\maketitle

The use of electric fields is a powerful approach to manipulate molecular electronic states\cite{Li2014,Kim2010,Wolkow_2005_Nature,Liu2021,Wan2019,Sahoo2005,PioroLadriere2008}, and consequently, optical properties, adsorption structures, vibrational frequencies, oxidation states and chemical reactivity \cite{Shaik2018,Gorin2012,Park2021,Alemani2006,Wolkow_2005_Nature,Croce2014,Kulzer1999,Mangel2020}. Being able to study these effects at the single molecule level would be very important for understanding the intimate interaction between molecules and their electrostatic environment.  Yet, performing such an experiment in a well-controlled manner has proven to be extremely difficult and scanning tunneling microscopy (STM) has emerged as a leading technique in this challenging field \cite{FernandezTorrente2012,Lee2018,Roslawska2022}. In STM, a significant electric field is present between the STM tip and the sample surface, which will induce a Stark shift of the electronic states observed in the tunneling spectra \cite{Limot2003,Kroeger2004}. By increasing the set-point tunneling current, the tip-sample distance decreases, leading to increasing electric field strength. Although this is a powerful experimental technique to study the effect of external electric fields on molecular electronic states, molecules are often required to be decoupled from a metallic substrate\cite{Repp2005,Qiu2003,Schulz2013}, due to the strong perturbation of their electronic states by hybridization, charge transfer, and screening with the metal substrate \cite{Tautz2007,Lu2004}. Finally, the tunneling current and electric field are linked and using high tunneling currents often leads to instabilities in the tip-molecule-sample junction.

\begin{figure}[t!]
    \centering
    \includegraphics[width = 1\textwidth]{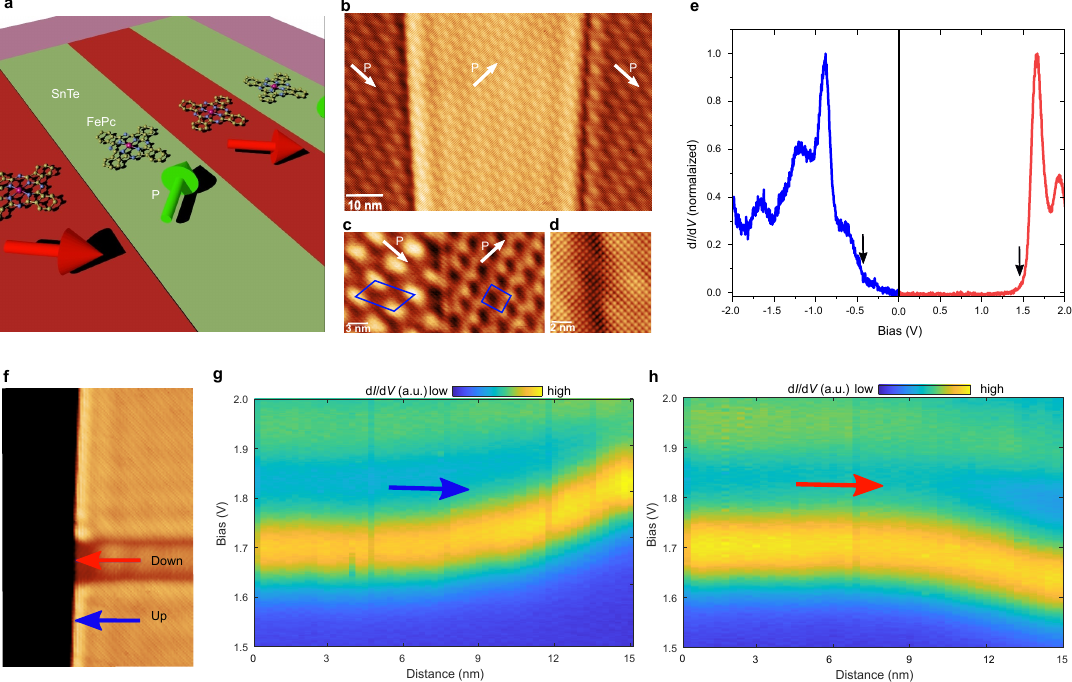}
    \caption{\textbf{a:} Schematics of the FePc molecules on SnTe (green and red area show the ferroelectric domains and the arrows show the direction of electric polarization in each domain). \textbf{b-d:} STM images of the domain formation , moir\'e pattern and atomic resolution of the SnTe monolayer, respectively (white arrows show the direction of the  electric polarization). \textbf{e:} Differential conductance (\dIdV) spectrum on SnTe (set point: $ I=1 \mathrm{~nA},V= -2\mathrm{~V}$ for blue spectrum and, $ I=1 \mathrm{~nA},V= 2 \mathrm{~V} $ for red spectrum). \textbf{f}: STM scan on SnTe shows bright and dark edges corresponding to band bending up (g) and down (h), respectively. \textbf{g-h}: Line spectra taken over $15 \mathrm{~nm}$ distance to the edge of the island as shown with blue and red arrows in panel f.}
\label{fig:SnTe}
\end{figure}

We overcome these limitations by coupling single molecules with two-dimensional ferroelectric (2D-FE) materials as shown schematically in Fig.~\ref{fig:SnTe}a. By controlling the charge polarization $\vec{P}$ of the FE, one can tune the electric field experienced by the molecules and consequently, their electronic states. This setup has the distinct advantage that the polarization direction of the FE substrate can be independently controlled and switched irrespective of the electric field from the STM tip. In addition, due to the semiconducting nature of the ferroelectric substrate, it effectively decouples the molecule from the metallic substrate, which gives access to the electronic states of essentially an isolated molecule. Here, we use a monolayer of tin telluride (SnTe) as our FE substrate (see Methods). It has two polarization states (P$\uparrow$ and P$\downarrow$) that are stable up to room temperature and that can be switched by an external electric field\cite{Chang2016}.


As a prototype system, we focus on iron-phthalocyanine (FePc) molecules adsorbed on a 2D-FE SnTe substrate (Fig.~\ref{fig:SnTe}a). FePc molecules have partially empty $d$ orbitals in the central metal atom that cause interesting magnetic properties \cite{doi:10.1038/s41598-017-01668-6,doi.org/10.1038/s41467-018-05163-y}. We use low-temperature STM and scanning tunnelling spectroscopy (STS) to study how the molecular states are affected by an in-plane electric field from the 2D-FE SnTe substrate. In particular, we show that the orbital filling and degeneracy of $d$ orbitals of a single FePc changes due to the presence of electric field from the SnTe substrate. This intriguing phenomenon stems from distinct metal $d$-orbital occupation caused by electron transfer and energy-level shift associated with the polarization switch of the SnTe monolayer (Fig.~\ref{fig:SnTe}a). Furthermore, it is possible to manipulate the molecular states by controlling the polarization of the FE domain using STM. Finally, we have compared our experimental results with density-functional theory (DFT) calculations, which further support the effects caused by in-plane electric fields on the FePc molecular states. Our study provides a well-defined, controllable platform for manipulation of molecular electronic states with an electric field, having also great potential for practical applications in molecular electronic and spintronic devices.

We first study the FE order of ultrathin SnTe monolayer grown by molecular beam epitaxy (MBE) on highly oriented pyrolytic graphite (HOPG) substrate (see Methods and Supporting Information Fig.~S1). Fig.~\ref{fig:SnTe}b shows an atomically resolved STM image of the SnTe monolayer with stripe domains, which are consistent with the domain structures observed on SnTe monolayer grown on a graphene substrate \cite{Chang2016}. 
The STM topography also exhibits a clear, well-ordered superstructure arising from the moir\'e pattern between the quasi-square SnTe lattice and hexagonal HOPG lattice. A detailed analysis of moir\'e pattern between SnTe lattice and hexagonal HOPG lattice can be found in the Supporting Information (SI) (see Fig.~S4 and S5 in the SI). As shown in Fig.~\ref{fig:SnTe}c, the domains with different polarization directions have different moir\'e unit cells due to the different distortion of the SnTe lattice. 
Finally, it is important to note that the lattice is continuous across the domain boundary as shown  in Fig.~\ref{fig:SnTe}d.
Fig.~\ref{fig:SnTe}e shows the typical differential conductance (\dIdV) spectra acquired on monolayer SnTe (in the middle of the domain). The \dIdV signal of conduction and valence bands has a large difference in intensity and we use different tunneling conditions for positive and negative bias (red and blue lines, respectively). The arrows in the \dIdV curve (Fig.~\ref{fig:SnTe}e) indicate the band edges giving a band gap of 1.85 eV.

As was shown previously\cite{Chang2016,Chang2020a,Zhang2022}, ferroelectric materials possess four characteristic features:
the formation of the domain structure, the presence of a lattice distortion and band-bending, and the possibility to manipulate the domain structure by external electric fields. 
As shown in Fig.~\ref{fig:SnTe}b and Fig.~S1a in SI, we have observed clear domain structure in our STM topography. Moreover, a detailed analysis of atomically resolved images further reveals that the lattice is slightly distorted from a perfect square to a parallelogram (see Fig. S1b in the SI). The signatures of band-bending can be observed by following the conduction band edges  at 1.7 V in the \dIdV curves as a function of the distance to a SnTe island edge seen in Fig.~\ref{fig:SnTe}f. Spatially resolved \dIdV spectra (Fig.~\ref{fig:SnTe}g and h) are taken along the lines perpendicular to the edges of two adjacent domains (blue and red arrows in Fig.~\ref{fig:SnTe}f). The conduction band onsets shift to opposite directions by up to 0.12 eV with a screening length of about 10 nm. Based on the band bending and the direction of lattice distortion, we can unambiguously determine the in-plane polarization direction (see Fig. S1b in the SI). 
Finally, we use a voltage pulse (4V) between the STM tip and the sample to successfully manipulate the FE polarization through domain wall motion (see Fig. S2 in the SI). The above observations uniquely demonstrate the existence of ferroelectricity in the system. 

\begin{figure}[t!]
    \centering
    \includegraphics[width =1 \textwidth]{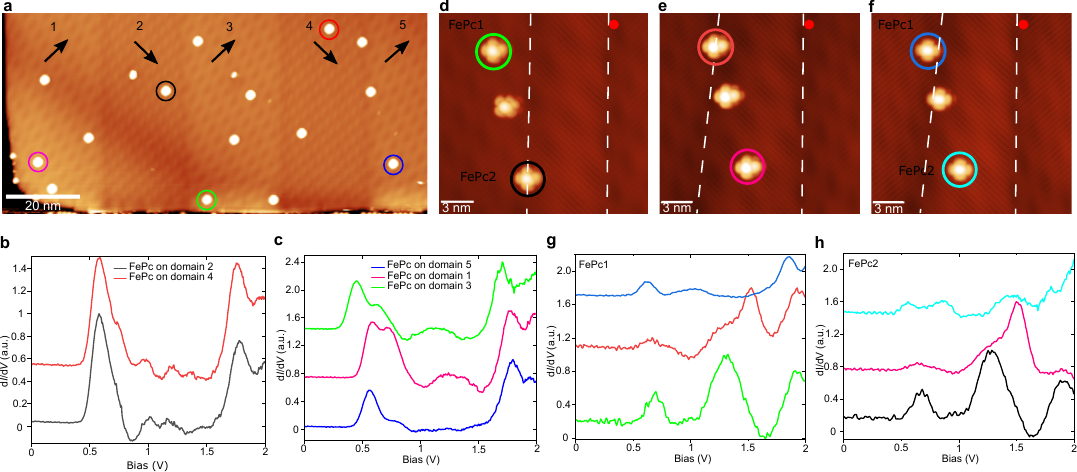}
    \caption{\textbf{a:} Atomically resolved STM image of the SnTe monolayer with FePc molecules (image size $ 120\times 55 \mathrm{~nm}^{2}, V= 2 \mathrm{~V}, I= 0.3 \mathrm{~nA}$). \textbf{b:} \dIdV spectra taken on FePcs sitting on domain 2 (indicated with black circle) and domain 4 (indicated with red circle). \textbf{c:} \dIdV spectra taken on FePcs sitting on domain 1 (indicated with pink circle), domain 3 (indicated with green circle) and domain 5 (indicated with blue circle). \textbf{d-f:} Manipulation of the ferroelectric domains using 4V bias voltage pulses applied at the position marked with the red dot. \textbf{g,h:} Point spectra taken on FePc 1 and 2 in panels d and f while manipulating the domains.  }\label{fig:SnTe with single molecles}
\end{figure}

Having demonstrated the ferroelectricity of monolayer SnTe by structural and spectroscopic measurements, we now turn to the coupling of this electronic order with magnetic molecular states in a single molecule. Fig.~\ref{fig:SnTe with single molecles}a shows topographic STM images of isolated FePc molecules that are adsorbed on different FE domains on SnTe. The direction of the polarization is indicated by arrows in Fig.~\ref{fig:SnTe with single molecles}a; these directions can be determined by lattice distortion together with the sign change of polarization charge on edges. Inspection of atomically resolved images demonstrates that FePc has two adsorption geometries which are rotated 45 degrees with respect to each other, with the central Fe atom either sitting on top of an Sn atom or on top of a Te atom of the underlying SnTe surface (see Fig.~S6 in the SI). DFT calculations confirm these two configurations as the most stable ones, and that both the adsorption site and the angle between the FePc molecule and the SnTe substrate play an important role for the stability of the system. The most energetically stable case is when the central atom Fe of the FePc molecule sits on top of Sn while one of the arms of the molecule (the line formed by two consecutive benzene rings) has an angle of $\theta = 45^{\circ}$ with one of the SnTe lattice vectors. The second most stable configuration (135 meV higher total energy) occurs when the central atom sits on top of Te while $\theta = 0^{\circ}$ (see Fig.~S5 in the SI). In both cases the FePc molecule keeps its planar geometry, and the Fe-Sn and Fe-Te distances are, respectively, 3.62 and 3.31 \AA. Additionally, the SnTe lattice parameters and its intrinsic polarization are not strongly affected by the presence of the molecule. The smaller lattice parameter of the $5\cross5$ SnTe supercell in both cases is 2.28 nm, while the largest parameter is $2.5\%$ larger.

\begin{figure}[t!]
    \centering
    \includegraphics[width = 1\textwidth]{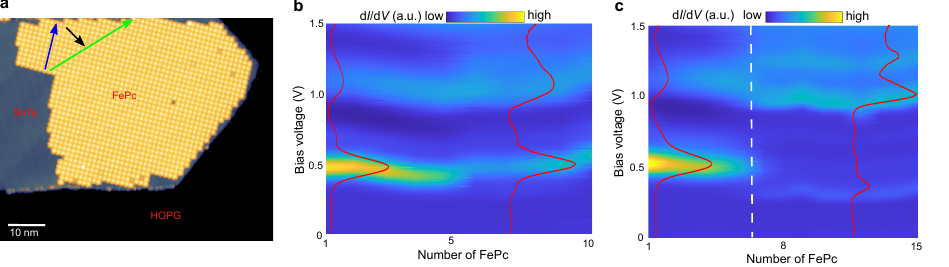}
    \caption{\textbf{a:} FePc island on SnTe (image size  $ 80\times 60 \mathrm{~nm}^{2}, V= 1.5 \mathrm{~V}, I= 300 \mathrm{~pA} $). \textbf{b:} Line spectra over a distance of 12.5 nm inside a single FE domain (blue arrow). {\textbf{c:} Line spectra over 27 nm crossing two domains (green arrow. Black arrow shows the position of boundary between two domains).}}
\label{fig:SnTe with island of molecles}
\end{figure}

Fig.~\ref{fig:SnTe with single molecles}b,c shows \dIdV point spectra taken on different molecules (spectra were obtained by positioning the tip over the central Fe atom and the molecule positions are marked with circles in Fig.~\ref{fig:SnTe with single molecles}a) with the same adsorption site and same orientation but located on different domains. 
As the direction of the ferroelectric polarization varies from domain to domain between two different values, molecules on domains 1, 3, and 5, and molecules on domains 2 and 4 feel the same polarization direction, respectively.  The spectrum obtained on the molecules show three main peaks at around  0.6, 1.2 and 1.8 V. The peaks at around 0.6 V and 1.2 V can be interpreted as resonances originating from the lowest unoccupied molecular orbital (LUMO) and the LUMO+1. The peak located at around 1.8 V corresponds to the SnTe conduction band, which also shifts slightly depending on the exact location where the spectra were measured.  Interestingly, the energy position and intensity of the LUMO and LUMO+1 resonances change depending on the polarization of FE domain. In particular, the single LUMO peak of FePc adsorbed on domain 2, 4 (Fig.~\ref{fig:SnTe with single molecles}b) splits when they are adsorbed on domain 1, 3, 5 (Fig.~\ref{fig:SnTe with single molecles}c). Furthermore, there are less intense features at around 1 V, which may come from further splitting of LUMO+1 peaks. We will discuss this in more detail in Fig.~\ref{fig:SnTe with island of molecles}.

In order to follow the relation between polarization of an FE domain and a change in the molecular states, we have performed a controlled ferroelectric domain manipulation by applying bias voltage pulses with an STM tip. The domain manipulation process is demonstrated as a series of pulses applied with the STM tip placed at the red dot shown in Fig.~\ref{fig:SnTe with single molecles}d-f. During the domain manipulation, the position of the top and bottom molecules is not changed, only the FE domain under the molecule is manipulated in a controlled manner. The corresponding \dIdV point spectra taken after each manipulation step are shown in Fig.~\ref{fig:SnTe with single molecles}g,h. Again, the change in the molecular states corresponds to the change of the domain manipulation. In particular, the energy position and intensity of the molecular resonances changes depend on the polarization of FE domain.

It is important to note that the voltage pulses also affect the final condition of the tip and hence the STS. In order to reliably show the effect of electric field caused by the ferroelectric layer on the molecular states, we have created an array of molecules. This allows us to investigate the molecular states as a function of position both within the same domain and across domains without an undesired changed of the STM tip apex. Fig.~\ref{fig:SnTe with island of molecles}a shows densely-packed islands of FePc molecules on top of SnTe (see Methods). This results in FePc islands that span multiple ferroelectric domains of the SnTe layer. We probed the effect of the ferroelectric domains on the \dIdV spectra by first measuring the molecular spectra over a single ferroelectric domain (blue arrow in Fig.~\ref{fig:SnTe with island of molecles}a). As we can see in Fig.~\ref{fig:SnTe with island of molecles}b, the spectra show some variations, but are qualitatively similar. There are two set of peaks at around 0.5 V and 1.2 V, which correspond to the LUMO and LUMO+1, respectively. The peak positions shift along the band bending of the SnTe layer. The effect of electric field on the molecules in the edge of the domain is stronger than on the molecules in the middle, as can be clearly seen from the last part of line spectra in Fig.~\ref{fig:SnTe with island of molecles}b.

We can also visualize the changes from one ferroelectric domain to another across the boundary between the domains. This is shown in the line spectra on the molecules along the green arrow. We observe that in crossing the boundary between two domains (black arrow in the Fig.~\ref{fig:SnTe with island of molecles}a), there is a discrete change in the \dIdV spectra (the FE domain boundary is indicated by the white dashed line in Fig.~\ref{fig:SnTe with island of molecles}c ). In particular, we observe that the original LUMO and LUMO+1 peaks split and intensities are inverted once the direction of polarization changes. This is consistent with our observations on single molecules discussed above. It is important to note that this splitting is not related to  where the molecules are located with respect to the underlying moir\'e pattern. In fact, the moir\'e pattern only periodically modulates the energy position of the conduction band of SnTe (see Fig. S4 in the SI). We have repeated the same experiment on different FePc islands and always observe the same behavior (see Fig. S6 in the SI). The main reason for this change is the Stark effect, which shifts and splits of molecular resonances due to the presence of an external electric field \cite{2018}. However, in our case, the electric field comes from the underlying FE substrate and it is not related to the electric field from the STM tip\cite{Lee2018, Roslawska2022}. Under this electric field, the $D_{4h}$  symmetry of FePc molecule is broken due to the coupling with ferroelectricity, and this further causes splitting of the partially occupied $d_{xz}$ and $d_{yz}$ levels of the FePc molecules as predicted by our DFT calculations (see detail below). 

\begin{figure}[t!]
    \centering
    \includegraphics[width = 1\textwidth]{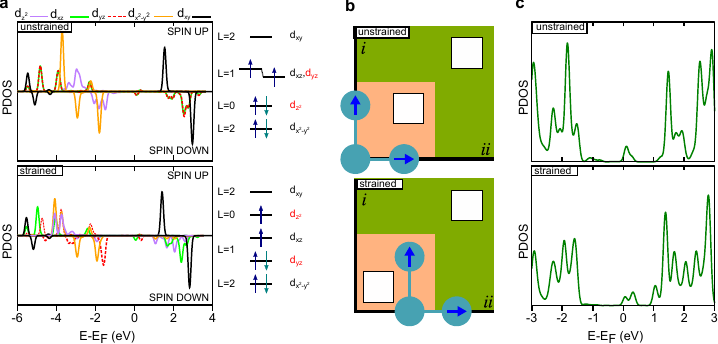}
    \caption{\textbf{a:} Projected density of states on the d orbitals of the Fe atom in the FePc molecule. The energy levels show the orbital transition caused by coupling with ferroelectricity. \textbf{b:} Schematic visualization of the importance of strain to the orbital transition. The arrows indicate the direction of ferroelectricity while the $i$ and $ii$ indices depict the direction of strain. The spin density around the Fe atom before and after the transition are shown as inset in the schematic image (isosurface 0.005 a.u.) \textbf{c:} Total density of states of the Fe atom in the FePc+SnTe system.}
\label{fig:DFT single molecule n SnTe}
\end{figure}

We have performed DFT calculations in order to understand the ground state and electronic properties of the FePc molecules in the presence of ferroelectricity from the SnTe substrate. The ground state of the isolated FePc molecule is a triplet $S=1$ state, with the spin polarization mostly concentrated on the central Fe atom, as has been predicted before through DFT and Monte Carlo simulations \cite{doi:10.1038/s41598-017-01668-6, doi.org/10.1038/s41467-018-05163-y, PhysRevB.85.155437}. The Fe 3$d$-electrons can also manifest the triplet state in different ways depending on the interaction with the substrate, and even a high spin quintuplet can be observed when FePc islands are deposited on a Cu surface \cite{PhysRevB.91.214427,doi:10.1063/1.4940138}. Here we demonstrate that the triplet can also correspond to different spin configurations depending on the coupling of the FePc molecule with the ferroelectricity of the SnTe substrate. In our calculations, the different domains of the SnTe layer observed in the experiment were simulated by first fully relaxing the FePc+SnTe system, giving a lattice distortion of 2.5 \%, and a second case was considered by increasing the distortion to 4\% to model a strained system. Fig. \ref{fig:DFT single molecule n SnTe}a show that the $D_{4h}$ tetragonal symmetry is slightly broken due to the coupling with ferroelectricity, and a subtle splitting of the partially occupied $d_{xz}$ and $d_{yz}$ levels is observed. Applying strain to the SnTe layer, hence increasing the coupling of the Fe states with the ferroelectricity, causes the promotion of one electron from the $d_{z^2}$ to the $d_{yz}$ orbital.
This transition driven by the combination of ferroelectricity and strain can be rationalized in terms of the low energy crystal field Hamiltonian of the molecule. In particular, given the symmetry of the system, the crystal field Hamiltonian for the Fe $d$-orbitals takes the form
\begin{equation}
    H = D l_z^2 + E (l_x^4 + l_y^4) + F l_z^4 + G (l_x^2-l_y^2)
\end{equation}
where $l_x, l_y, l_z$ are the single particle angular momentum operators in the Fe $d$-manifold. The physical significance of the different terms can be understood as follows. The terms $D$ and $F$ account for the planar nature of the molecule,
$E$ for the four-fold rotational symmetry and $G$ controls the induced breaking of rotational symmetry induced by the
ferroelectric strained substrate.
We first note that in the absence of strain in the sample, the two directions of the ferroelectric
polarization would be equivalent due to the original $C_4$ symmetry of the substrate. In this scenario,
ferroelectric polarizations rotated by 90$^\circ$ must give rise to equivalent spectra due to symmetry, as depicted schematically in Fig.~\ref{fig:DFT single molecule n SnTe}b.
In contrast, in the presence of strain in the sample, two configurations with ferroelectric polarization
rotated by 90$^\circ$ will give rise to inequivalent electronic configurations, due to the explicit
breaking of $C_4$ created by the strain. In this scenario, the two ferroelectric configurations
will induce different values of $|G|$ in the molecule, effectively allowing to control its
state by the polarization of the underlying substrate. 
For small values of $|G|$, the crystal field gives rise to a spin density located in the
$d_{xz}$ and $d_{yz}$ orbitals. Once the induced breaking driven by the strain ferroelectric surpasses a critical value,
the term $G$ drives an orbital ordering transition yielding a spin polarization located in the
$d_{z^z}$ and $d_{xz}$ orbitals. The schematic image in Fig.~\ref{fig:DFT single molecule n SnTe}b shows the spin densities obtained through DFT around the Fe atom before and after the FePc undergoes the orbital transition. In particular, the symmetry breaking induced by $G$ drives a splitting between
the originally degenerate levels $d_{xz}$ and $d_{yz}$, accounting for the orbital transition in the molecule. Fig.~\ref{fig:DFT single molecule n SnTe}c shows that the orbital transition changes considerably the density of states of the Fe atom, mostly because the $d_{z^2}$ orbital is partially occupied after the transition, explaining the different d$I$/d$V$ spectra obtained when the FePc molecule is deposited in different domains. Fig.~S8 in the SI shows that the density of states projected on all atoms of the FePc molecule is strongly affected by the coupling with ferroelectricity as well.

In conclusion, we have proposed a new platform for probing the effect of an electric field on molecular orbitals by coupling single molecules with a two-dimensional ferroelectric material, with the possibility to manipulate the molecular states by controlling the polarization of the FE domains. In particular, we have demonstrated that under the presence of an intrinsic electric field  from the underlying FE substrate, the orbital filling and degeneracy of $d$ orbitals of a single FePc changes. This provides a promising way to achieve nonvolatile switching of magnetism at the molecular scale by a 2D ferroelectric substrate and has great potential for practical applications in logic and spintronics devices. As we control the magnetism in a single molecule through the FE polarization, it is also a first step towards constructing artifical multiferroic states in molecule-2D material-hybrids.


\bibliography{SnTe1}

\begin{thebibliography}{38}%
\makeatletter
\providecommand \@ifxundefined [1]{%
 \@ifx{#1\undefined}
}%
\providecommand \@ifnum [1]{%
 \ifnum #1\expandafter \@firstoftwo
 \else \expandafter \@secondoftwo
 \fi
}%
\providecommand \@ifx [1]{%
 \ifx #1\expandafter \@firstoftwo
 \else \expandafter \@secondoftwo
 \fi
}%
\providecommand \natexlab [1]{#1}%
\providecommand \enquote  [1]{``#1''}%
\providecommand \bibnamefont  [1]{#1}%
\providecommand \bibfnamefont [1]{#1}%
\providecommand \citenamefont [1]{#1}%
\providecommand \href@noop [0]{\@secondoftwo}%
\providecommand \href [0]{\begingroup \@sanitize@url \@href}%
\providecommand \@href[1]{\@@startlink{#1}\@@href}%
\providecommand \@@href[1]{\endgroup#1\@@endlink}%
\providecommand \@sanitize@url [0]{\catcode `\\12\catcode `\$12\catcode
  `\&12\catcode `\#12\catcode `\^12\catcode `\_12\catcode `\%12\relax}%
\providecommand \@@startlink[1]{}%
\providecommand \@@endlink[0]{}%
\providecommand \url  [0]{\begingroup\@sanitize@url \@url }%
\providecommand \@url [1]{\endgroup\@href {#1}{\urlprefix }}%
\providecommand \urlprefix  [0]{URL }%
\providecommand \Eprint [0]{\href }%
\providecommand \doibase [0]{http://dx.doi.org/}%
\providecommand \selectlanguage [0]{\@gobble}%
\providecommand \bibinfo  [0]{\@secondoftwo}%
\providecommand \bibfield  [0]{\@secondoftwo}%
\providecommand \translation [1]{[#1]}%
\providecommand \BibitemOpen [0]{}%
\providecommand \bibitemStop [0]{}%
\providecommand \bibitemNoStop [0]{.\EOS\space}%
\providecommand \EOS [0]{\spacefactor3000\relax}%
\providecommand \BibitemShut  [1]{\csname bibitem#1\endcsname}%
\let\auto@bib@innerbib\@empty
\bibitem [{\citenamefont {Li}\ \emph {et~al.}(2014)\citenamefont {Li},
  \citenamefont {Doak}, \citenamefont {Kronik}, \citenamefont {Neaton},\ and\
  \citenamefont {Natelson}}]{Li2014}%
  \BibitemOpen
  \bibfield  {author} {\bibinfo {author} {\bibfnamefont {Yajing}\ \bibnamefont
  {Li}}, \bibinfo {author} {\bibfnamefont {Peter}\ \bibnamefont {Doak}},
  \bibinfo {author} {\bibfnamefont {Leeor}\ \bibnamefont {Kronik}}, \bibinfo
  {author} {\bibfnamefont {Jeffrey~B.}\ \bibnamefont {Neaton}}, \ and\ \bibinfo
  {author} {\bibfnamefont {Douglas}\ \bibnamefont {Natelson}},\ }\bibfield
  {title} {\enquote {\bibinfo {title} {Voltage tuning of vibrational mode
  energies in single-molecule junctions},}\ }\href {\doibase
  10.1073/pnas.1320210111} {\bibfield  {journal} {\bibinfo  {journal} {Proc.
  Nat. Acad. Sci.}\ }\textbf {\bibinfo {volume} {111}},\ \bibinfo {pages}
  {1282--1287} (\bibinfo {year} {2014})}\BibitemShut {NoStop}%
\bibitem [{\citenamefont {Kim}\ and\ \citenamefont {Kim}(2010)}]{Kim2010}%
  \BibitemOpen
  \bibfield  {author} {\bibinfo {author} {\bibfnamefont {Woo~Youn}\
  \bibnamefont {Kim}}\ and\ \bibinfo {author} {\bibfnamefont {Kwang~S.}\
  \bibnamefont {Kim}},\ }\bibfield  {title} {\enquote {\bibinfo {title} {Tuning
  molecular orbitals in molecular electronics and spintronics},}\ }\href
  {\doibase 10.1021/ar900156u} {\bibfield  {journal} {\bibinfo  {journal} {Acc.
  Chem. Res.}\ }\textbf {\bibinfo {volume} {43}},\ \bibinfo {pages} {111--120}
  (\bibinfo {year} {2010})}\BibitemShut {NoStop}%
\bibitem [{\citenamefont {{Piva}}\ \emph {et~al.}(2005)\citenamefont {{Piva}},
  \citenamefont {{Dilabio}}, \citenamefont {{Pitters}}, \citenamefont
  {{Zikovsky}}, \citenamefont {{Rezeq}}, \citenamefont {{Dogel}}, \citenamefont
  {{Hofer}},\ and\ \citenamefont {{Wolkow}}}]{Wolkow_2005_Nature}%
  \BibitemOpen
  \bibfield  {author} {\bibinfo {author} {\bibfnamefont {Paul~G.}\ \bibnamefont
  {{Piva}}}, \bibinfo {author} {\bibfnamefont {Gino~A.}\ \bibnamefont
  {{Dilabio}}}, \bibinfo {author} {\bibfnamefont {Jason~L.}\ \bibnamefont
  {{Pitters}}}, \bibinfo {author} {\bibfnamefont {Janik}\ \bibnamefont
  {{Zikovsky}}}, \bibinfo {author} {\bibfnamefont {Moh'd.}\ \bibnamefont
  {{Rezeq}}}, \bibinfo {author} {\bibfnamefont {Stanislav}\ \bibnamefont
  {{Dogel}}}, \bibinfo {author} {\bibfnamefont {Werner~A.}\ \bibnamefont
  {{Hofer}}}, \ and\ \bibinfo {author} {\bibfnamefont {Robert~A.}\ \bibnamefont
  {{Wolkow}}},\ }\bibfield  {title} {\enquote {\bibinfo {title} {{Field
  regulation of single-molecule conductivity by a charged surface atom}},}\
  }\href {\doibase 10.1038/nature03563} {\bibfield  {journal} {\bibinfo
  {journal} {\nat}\ }\textbf {\bibinfo {volume} {435}},\ \bibinfo {pages}
  {658--661} (\bibinfo {year} {2005})}\BibitemShut {NoStop}%
\bibitem [{\citenamefont {Liu}\ \emph {et~al.}(2021)\citenamefont {Liu},
  \citenamefont {Laguta}, \citenamefont {Inzani}, \citenamefont {Huang},
  \citenamefont {Das}, \citenamefont {Chatterjee}, \citenamefont {Sheridan},
  \citenamefont {Griffin}, \citenamefont {Ardavan},\ and\ \citenamefont
  {Ramesh}}]{Liu2021}%
  \BibitemOpen
  \bibfield  {author} {\bibinfo {author} {\bibfnamefont {Junjie}\ \bibnamefont
  {Liu}}, \bibinfo {author} {\bibfnamefont {Valentin~V.}\ \bibnamefont
  {Laguta}}, \bibinfo {author} {\bibfnamefont {Katherine}\ \bibnamefont
  {Inzani}}, \bibinfo {author} {\bibfnamefont {Weichuan}\ \bibnamefont
  {Huang}}, \bibinfo {author} {\bibfnamefont {Sujit}\ \bibnamefont {Das}},
  \bibinfo {author} {\bibfnamefont {Ruchira}\ \bibnamefont {Chatterjee}},
  \bibinfo {author} {\bibfnamefont {Evan}\ \bibnamefont {Sheridan}}, \bibinfo
  {author} {\bibfnamefont {Sinéad~M.}\ \bibnamefont {Griffin}}, \bibinfo
  {author} {\bibfnamefont {Arzhang}\ \bibnamefont {Ardavan}}, \ and\ \bibinfo
  {author} {\bibfnamefont {Ramamoorthy}\ \bibnamefont {Ramesh}},\ }\bibfield
  {title} {\enquote {\bibinfo {title} {Coherent electric field manipulation of
  {Fe$^{3+}$ spins in PbTiO$_3$}},}\ }\href {\doibase 10.1126/sciadv.abf8103}
  {\bibfield  {journal} {\bibinfo  {journal} {Sci. Adv.}\ }\textbf {\bibinfo
  {volume} {7}},\ \bibinfo {pages} {eabf8103} (\bibinfo {year}
  {2021})}\BibitemShut {NoStop}%
\bibitem [{\citenamefont {Wan}\ \emph {et~al.}(2019)\citenamefont {Wan},
  \citenamefont {Li}, \citenamefont {Li}, \citenamefont {Mao}, \citenamefont
  {Wang}, \citenamefont {Chen}, \citenamefont {Dong}, \citenamefont {Nie},
  \citenamefont {Xiang}, \citenamefont {Liu}, \citenamefont {Zhu},\ and\
  \citenamefont {Zeng}}]{Wan2019}%
  \BibitemOpen
  \bibfield  {author} {\bibinfo {author} {\bibfnamefont {Siyuan}\ \bibnamefont
  {Wan}}, \bibinfo {author} {\bibfnamefont {Yue}\ \bibnamefont {Li}}, \bibinfo
  {author} {\bibfnamefont {Wei}\ \bibnamefont {Li}}, \bibinfo {author}
  {\bibfnamefont {Xiaoyu}\ \bibnamefont {Mao}}, \bibinfo {author}
  {\bibfnamefont {Chen}\ \bibnamefont {Wang}}, \bibinfo {author} {\bibfnamefont
  {Chen}\ \bibnamefont {Chen}}, \bibinfo {author} {\bibfnamefont {Jiyu}\
  \bibnamefont {Dong}}, \bibinfo {author} {\bibfnamefont {Anmin}\ \bibnamefont
  {Nie}}, \bibinfo {author} {\bibfnamefont {Jianyong}\ \bibnamefont {Xiang}},
  \bibinfo {author} {\bibfnamefont {Zhongyuan}\ \bibnamefont {Liu}}, \bibinfo
  {author} {\bibfnamefont {Wenguang}\ \bibnamefont {Zhu}}, \ and\ \bibinfo
  {author} {\bibfnamefont {Hualing}\ \bibnamefont {Zeng}},\ }\bibfield  {title}
  {\enquote {\bibinfo {title} {Nonvolatile ferroelectric memory effect in
  ultrathin {$\alpha$-In$_2$Se$_3$}},}\ }\href {\doibase
  https://doi.org/10.1002/adfm.201808606} {\bibfield  {journal} {\bibinfo
  {journal} {Adv. Funct. Mater.}\ }\textbf {\bibinfo {volume} {29}},\ \bibinfo
  {pages} {1808606} (\bibinfo {year} {2019})}\BibitemShut {NoStop}%
\bibitem [{\citenamefont {Sahoo}\ \emph {et~al.}(2005)\citenamefont {Sahoo},
  \citenamefont {Kontos}, \citenamefont {Furer}, \citenamefont {Hoffmann},
  \citenamefont {Gräber}, \citenamefont {Cottet},\ and\ \citenamefont
  {Schönenberger}}]{Sahoo2005}%
  \BibitemOpen
  \bibfield  {author} {\bibinfo {author} {\bibfnamefont {Sangeeta}\
  \bibnamefont {Sahoo}}, \bibinfo {author} {\bibfnamefont {Takis}\ \bibnamefont
  {Kontos}}, \bibinfo {author} {\bibfnamefont {Jürg}\ \bibnamefont {Furer}},
  \bibinfo {author} {\bibfnamefont {Christian}\ \bibnamefont {Hoffmann}},
  \bibinfo {author} {\bibfnamefont {Matthias}\ \bibnamefont {Gräber}},
  \bibinfo {author} {\bibfnamefont {Audrey}\ \bibnamefont {Cottet}}, \ and\
  \bibinfo {author} {\bibfnamefont {Christian}\ \bibnamefont
  {Schönenberger}},\ }\bibfield  {title} {\enquote {\bibinfo {title} {Electric
  field control of spin transport},}\ }\href {\doibase 10.1038/nphys149}
  {\bibfield  {journal} {\bibinfo  {journal} {Nat. Phys.}\ }\textbf {\bibinfo
  {volume} {1}},\ \bibinfo {pages} {99--102} (\bibinfo {year}
  {2005})}\BibitemShut {NoStop}%
\bibitem [{\citenamefont {Pioro-Ladrière}\ \emph {et~al.}(2008)\citenamefont
  {Pioro-Ladrière}, \citenamefont {Obata}, \citenamefont {Tokura},
  \citenamefont {Shin}, \citenamefont {Kubo}, \citenamefont {Yoshida},
  \citenamefont {Taniyama},\ and\ \citenamefont {Tarucha}}]{PioroLadriere2008}%
  \BibitemOpen
  \bibfield  {author} {\bibinfo {author} {\bibfnamefont {M.}~\bibnamefont
  {Pioro-Ladrière}}, \bibinfo {author} {\bibfnamefont {T.}~\bibnamefont
  {Obata}}, \bibinfo {author} {\bibfnamefont {Y.}~\bibnamefont {Tokura}},
  \bibinfo {author} {\bibfnamefont {Y.-S.}\ \bibnamefont {Shin}}, \bibinfo
  {author} {\bibfnamefont {T.}~\bibnamefont {Kubo}}, \bibinfo {author}
  {\bibfnamefont {K.}~\bibnamefont {Yoshida}}, \bibinfo {author} {\bibfnamefont
  {T.}~\bibnamefont {Taniyama}}, \ and\ \bibinfo {author} {\bibfnamefont
  {S.}~\bibnamefont {Tarucha}},\ }\bibfield  {title} {\enquote {\bibinfo
  {title} {Electrically driven single-electron spin resonance in a slanting
  {Zeeman} field},}\ }\href {\doibase 10.1038/nphys1053} {\bibfield  {journal}
  {\bibinfo  {journal} {Nat. Phys.}\ }\textbf {\bibinfo {volume} {4}},\
  \bibinfo {pages} {776--779} (\bibinfo {year} {2008})}\BibitemShut {NoStop}%
\bibitem [{\citenamefont {Shaik}\ \emph {et~al.}(2018)\citenamefont {Shaik},
  \citenamefont {Ramanan}, \citenamefont {Danovich},\ and\ \citenamefont
  {Mandal}}]{Shaik2018}%
  \BibitemOpen
  \bibfield  {author} {\bibinfo {author} {\bibfnamefont {Sason}\ \bibnamefont
  {Shaik}}, \bibinfo {author} {\bibfnamefont {Rajeev}\ \bibnamefont {Ramanan}},
  \bibinfo {author} {\bibfnamefont {David}\ \bibnamefont {Danovich}}, \ and\
  \bibinfo {author} {\bibfnamefont {Debasish}\ \bibnamefont {Mandal}},\
  }\bibfield  {title} {\enquote {\bibinfo {title} {Structure and
  reactivity/selectivity control by oriented-external electric fields},}\
  }\href {\doibase 10.1039/C8CS00354H} {\bibfield  {journal} {\bibinfo
  {journal} {Chem. Soc. Rev.}\ }\textbf {\bibinfo {volume} {47}},\ \bibinfo
  {pages} {5125--5145} (\bibinfo {year} {2018})}\BibitemShut {NoStop}%
\bibitem [{\citenamefont {Gorin}\ \emph {et~al.}(2012)\citenamefont {Gorin},
  \citenamefont {Beh},\ and\ \citenamefont {Kanan}}]{Gorin2012}%
  \BibitemOpen
  \bibfield  {author} {\bibinfo {author} {\bibfnamefont {Craig~F.}\
  \bibnamefont {Gorin}}, \bibinfo {author} {\bibfnamefont {Eugene~S.}\
  \bibnamefont {Beh}}, \ and\ \bibinfo {author} {\bibfnamefont {Matthew~W.}\
  \bibnamefont {Kanan}},\ }\bibfield  {title} {\enquote {\bibinfo {title} {An
  electric field–induced change in the selectivity of a metal
  oxide–catalyzed epoxide rearrangement},}\ }\href {\doibase
  10.1021/ja210365j} {\bibfield  {journal} {\bibinfo  {journal} {J. Am. Chem.
  Soc.}\ }\textbf {\bibinfo {volume} {134}},\ \bibinfo {pages} {186--189}
  (\bibinfo {year} {2012})}\BibitemShut {NoStop}%
\bibitem [{\citenamefont {Park}\ \emph {et~al.}(2021)\citenamefont {Park},
  \citenamefont {Shin},\ and\ \citenamefont {Kang}}]{Park2021}%
  \BibitemOpen
  \bibfield  {author} {\bibinfo {author} {\bibfnamefont {Youngwook}\
  \bibnamefont {Park}}, \bibinfo {author} {\bibfnamefont {Sunghwan}\
  \bibnamefont {Shin}}, \ and\ \bibinfo {author} {\bibfnamefont {Heon}\
  \bibnamefont {Kang}},\ }\bibfield  {title} {\enquote {\bibinfo {title}
  {Recent progress in the manipulation of molecules with dc electric fields},}\
  }\href {\doibase 10.1021/acs.accounts.0c00609} {\bibfield  {journal}
  {\bibinfo  {journal} {Acc. Chem. Res.}\ }\textbf {\bibinfo {volume} {54}},\
  \bibinfo {pages} {323--331} (\bibinfo {year} {2021})}\BibitemShut {NoStop}%
\bibitem [{\citenamefont {Alemani}\ \emph {et~al.}(2006)\citenamefont
  {Alemani}, \citenamefont {Peters}, \citenamefont {Hecht}, \citenamefont
  {Rieder}, \citenamefont {Moresco},\ and\ \citenamefont
  {Grill}}]{Alemani2006}%
  \BibitemOpen
  \bibfield  {author} {\bibinfo {author} {\bibfnamefont {Micol}\ \bibnamefont
  {Alemani}}, \bibinfo {author} {\bibfnamefont {Maike~V.}\ \bibnamefont
  {Peters}}, \bibinfo {author} {\bibfnamefont {Stefan}\ \bibnamefont {Hecht}},
  \bibinfo {author} {\bibfnamefont {Karl-Heinz}\ \bibnamefont {Rieder}},
  \bibinfo {author} {\bibfnamefont {Francesca}\ \bibnamefont {Moresco}}, \ and\
  \bibinfo {author} {\bibfnamefont {Leonhard}\ \bibnamefont {Grill}},\
  }\bibfield  {title} {\enquote {\bibinfo {title} {Electric field-induced
  isomerization of azobenzene by {STM}},}\ }\href {\doibase 10.1021/ja065449s}
  {\bibfield  {journal} {\bibinfo  {journal} {J. Am. Chem. Soc.}\ }\textbf
  {\bibinfo {volume} {128}},\ \bibinfo {pages} {14446--14447} (\bibinfo {year}
  {2006})}\BibitemShut {NoStop}%
\bibitem [{\citenamefont {Croce}\ and\ \citenamefont {van
  Amerongen}(2014)}]{Croce2014}%
  \BibitemOpen
  \bibfield  {author} {\bibinfo {author} {\bibfnamefont {Roberta}\ \bibnamefont
  {Croce}}\ and\ \bibinfo {author} {\bibfnamefont {Herbert}\ \bibnamefont {van
  Amerongen}},\ }\bibfield  {title} {\enquote {\bibinfo {title} {Natural
  strategies for photosynthetic light harvesting},}\ }\href {\doibase
  10.1038/nchembio.1555} {\bibfield  {journal} {\bibinfo  {journal} {Nat. Chem.
  Biol.}\ }\textbf {\bibinfo {volume} {10}},\ \bibinfo {pages} {492--501}
  (\bibinfo {year} {2014})}\BibitemShut {NoStop}%
\bibitem [{\citenamefont {Kulzer}\ \emph {et~al.}(1999)\citenamefont {Kulzer},
  \citenamefont {Matzke}, \citenamefont {Bräuchle},\ and\ \citenamefont
  {Basché}}]{Kulzer1999}%
  \BibitemOpen
  \bibfield  {author} {\bibinfo {author} {\bibfnamefont {F.}~\bibnamefont
  {Kulzer}}, \bibinfo {author} {\bibfnamefont {R.}~\bibnamefont {Matzke}},
  \bibinfo {author} {\bibfnamefont {C.}~\bibnamefont {Bräuchle}}, \ and\
  \bibinfo {author} {\bibfnamefont {Th.}\ \bibnamefont {Basché}},\ }\bibfield
  {title} {\enquote {\bibinfo {title} {Nonphotochemical hole burning
  investigated at the single-molecule level: Stark effect measurements on the
  original and photoproduct state},}\ }\href {\doibase 10.1021/jp9839448}
  {\bibfield  {journal} {\bibinfo  {journal} {J. Phys. Chem. A}\ }\textbf
  {\bibinfo {volume} {103}},\ \bibinfo {pages} {2408--2411} (\bibinfo {year}
  {1999})}\BibitemShut {NoStop}%
\bibitem [{\citenamefont {Mangel}\ \emph {et~al.}(2020)\citenamefont {Mangel},
  \citenamefont {Skripnik}, \citenamefont {Polyudov}, \citenamefont {Dette},
  \citenamefont {Wollandt}, \citenamefont {Punke}, \citenamefont {Li},
  \citenamefont {Urcuyo}, \citenamefont {Pauly}, \citenamefont {Jung},\ and\
  \citenamefont {Kern}}]{Mangel2020}%
  \BibitemOpen
  \bibfield  {author} {\bibinfo {author} {\bibfnamefont {Shai}\ \bibnamefont
  {Mangel}}, \bibinfo {author} {\bibfnamefont {Maxim}\ \bibnamefont
  {Skripnik}}, \bibinfo {author} {\bibfnamefont {Katharina}\ \bibnamefont
  {Polyudov}}, \bibinfo {author} {\bibfnamefont {Christian}\ \bibnamefont
  {Dette}}, \bibinfo {author} {\bibfnamefont {Tobias}\ \bibnamefont
  {Wollandt}}, \bibinfo {author} {\bibfnamefont {Paul}\ \bibnamefont {Punke}},
  \bibinfo {author} {\bibfnamefont {Dongzhe}\ \bibnamefont {Li}}, \bibinfo
  {author} {\bibfnamefont {Roberto}\ \bibnamefont {Urcuyo}}, \bibinfo {author}
  {\bibfnamefont {Fabian}\ \bibnamefont {Pauly}}, \bibinfo {author}
  {\bibfnamefont {Soon~Jung}\ \bibnamefont {Jung}}, \ and\ \bibinfo {author}
  {\bibfnamefont {Klaus}\ \bibnamefont {Kern}},\ }\bibfield  {title} {\enquote
  {\bibinfo {title} {Electric-field control of single-molecule
  tautomerization},}\ }\href {\doibase 10.1039/C9CP06868F} {\bibfield
  {journal} {\bibinfo  {journal} {Phys. Chem. Chem. Phys.}\ }\textbf {\bibinfo
  {volume} {22}},\ \bibinfo {pages} {6370--6375} (\bibinfo {year}
  {2020})}\BibitemShut {NoStop}%
\bibitem [{\citenamefont {Fern\'andez-Torrente}\ \emph
  {et~al.}(2012)\citenamefont {Fern\'andez-Torrente}, \citenamefont
  {Kreikemeyer-Lorenzo}, \citenamefont {Str\'o\ifmmode~\dot{z}\else
  \.{z}\fi{}ecka}, \citenamefont {Franke},\ and\ \citenamefont
  {Pascual}}]{FernandezTorrente2012}%
  \BibitemOpen
  \bibfield  {author} {\bibinfo {author} {\bibfnamefont {I.}~\bibnamefont
  {Fern\'andez-Torrente}}, \bibinfo {author} {\bibfnamefont {D.}~\bibnamefont
  {Kreikemeyer-Lorenzo}}, \bibinfo {author} {\bibfnamefont {A.}~\bibnamefont
  {Str\'o\ifmmode~\dot{z}\else \.{z}\fi{}ecka}}, \bibinfo {author}
  {\bibfnamefont {K.~J.}\ \bibnamefont {Franke}}, \ and\ \bibinfo {author}
  {\bibfnamefont {J.~I.}\ \bibnamefont {Pascual}},\ }\bibfield  {title}
  {\enquote {\bibinfo {title} {Gating the charge state of single molecules by
  local electric fields},}\ }\href {\doibase 10.1103/PhysRevLett.108.036801}
  {\bibfield  {journal} {\bibinfo  {journal} {Phys. Rev. Lett.}\ }\textbf
  {\bibinfo {volume} {108}},\ \bibinfo {pages} {036801} (\bibinfo {year}
  {2012})}\BibitemShut {NoStop}%
\bibitem [{\citenamefont {Lee}\ \emph {et~al.}(2018)\citenamefont {Lee},
  \citenamefont {Tallarida}, \citenamefont {Chen}, \citenamefont {Jensen},\
  and\ \citenamefont {Apkarian}}]{Lee2018}%
  \BibitemOpen
  \bibfield  {author} {\bibinfo {author} {\bibfnamefont {Joonhee}\ \bibnamefont
  {Lee}}, \bibinfo {author} {\bibfnamefont {Nicholas}\ \bibnamefont
  {Tallarida}}, \bibinfo {author} {\bibfnamefont {Xing}\ \bibnamefont {Chen}},
  \bibinfo {author} {\bibfnamefont {Lasse}\ \bibnamefont {Jensen}}, \ and\
  \bibinfo {author} {\bibfnamefont {V.~Ara}\ \bibnamefont {Apkarian}},\
  }\bibfield  {title} {\enquote {\bibinfo {title} {Microscopy with a
  single-molecule scanning electrometer},}\ }\href {\doibase
  10.1126/sciadv.aat5472} {\bibfield  {journal} {\bibinfo  {journal} {Sci.
  Adv.}\ }\textbf {\bibinfo {volume} {4}},\ \bibinfo {pages} {eaat5472}
  (\bibinfo {year} {2018})}\BibitemShut {NoStop}%
\bibitem [{\citenamefont {Ros\l{}awska}\ \emph {et~al.}(2022)\citenamefont
  {Ros\l{}awska}, \citenamefont {Neuman}, \citenamefont {Doppagne},
  \citenamefont {Borisov}, \citenamefont {Romeo}, \citenamefont {Scheurer},
  \citenamefont {Aizpurua},\ and\ \citenamefont {Schull}}]{Roslawska2022}%
  \BibitemOpen
  \bibfield  {author} {\bibinfo {author} {\bibfnamefont {Anna}\ \bibnamefont
  {Ros\l{}awska}}, \bibinfo {author} {\bibfnamefont {Tom\'a\ifmmode
  \check{s}\else~\v{s}\fi{}}\ \bibnamefont {Neuman}}, \bibinfo {author}
  {\bibfnamefont {Benjamin}\ \bibnamefont {Doppagne}}, \bibinfo {author}
  {\bibfnamefont {Andrei~G.}\ \bibnamefont {Borisov}}, \bibinfo {author}
  {\bibfnamefont {Michelangelo}\ \bibnamefont {Romeo}}, \bibinfo {author}
  {\bibfnamefont {Fabrice}\ \bibnamefont {Scheurer}}, \bibinfo {author}
  {\bibfnamefont {Javier}\ \bibnamefont {Aizpurua}}, \ and\ \bibinfo {author}
  {\bibfnamefont {Guillaume}\ \bibnamefont {Schull}},\ }\bibfield  {title}
  {\enquote {\bibinfo {title} {Mapping {Lamb}, {Stark}, and {Purcell} effects
  at a chromophore-picocavity junction with hyper-resolved fluorescence
  microscopy},}\ }\href {\doibase 10.1103/PhysRevX.12.011012} {\bibfield
  {journal} {\bibinfo  {journal} {Phys. Rev. X}\ }\textbf {\bibinfo {volume}
  {12}},\ \bibinfo {pages} {011012} (\bibinfo {year} {2022})}\BibitemShut
  {NoStop}%
\bibitem [{\citenamefont {Limot}\ \emph {et~al.}(2003)\citenamefont {Limot},
  \citenamefont {Maroutian}, \citenamefont {Johansson},\ and\ \citenamefont
  {Berndt}}]{Limot2003}%
  \BibitemOpen
  \bibfield  {author} {\bibinfo {author} {\bibfnamefont {L.}~\bibnamefont
  {Limot}}, \bibinfo {author} {\bibfnamefont {T.}~\bibnamefont {Maroutian}},
  \bibinfo {author} {\bibfnamefont {P.}~\bibnamefont {Johansson}}, \ and\
  \bibinfo {author} {\bibfnamefont {R.}~\bibnamefont {Berndt}},\ }\bibfield
  {title} {\enquote {\bibinfo {title} {Surface-state {Stark} shift in a
  scanning tunneling microscope},}\ }\href {\doibase
  10.1103/PhysRevLett.91.196801} {\bibfield  {journal} {\bibinfo  {journal}
  {Phys. Rev. Lett.}\ }\textbf {\bibinfo {volume} {91}},\ \bibinfo {pages}
  {196801} (\bibinfo {year} {2003})}\BibitemShut {NoStop}%
\bibitem [{\citenamefont {Kr\"oger}\ \emph {et~al.}(2004)\citenamefont
  {Kr\"oger}, \citenamefont {Limot}, \citenamefont {Jensen}, \citenamefont
  {Berndt},\ and\ \citenamefont {Johansson}}]{Kroeger2004}%
  \BibitemOpen
  \bibfield  {author} {\bibinfo {author} {\bibfnamefont {J.}~\bibnamefont
  {Kr\"oger}}, \bibinfo {author} {\bibfnamefont {L.}~\bibnamefont {Limot}},
  \bibinfo {author} {\bibfnamefont {H.}~\bibnamefont {Jensen}}, \bibinfo
  {author} {\bibfnamefont {R.}~\bibnamefont {Berndt}}, \ and\ \bibinfo {author}
  {\bibfnamefont {P.}~\bibnamefont {Johansson}},\ }\bibfield  {title} {\enquote
  {\bibinfo {title} {Stark effect in $\text{Au}(111)$ and $\text{Cu}(111)$
  surface states},}\ }\href {\doibase 10.1103/PhysRevB.70.033401} {\bibfield
  {journal} {\bibinfo  {journal} {Phys. Rev. B}\ }\textbf {\bibinfo {volume}
  {70}},\ \bibinfo {pages} {033401} (\bibinfo {year} {2004})}\BibitemShut
  {NoStop}%
\bibitem [{\citenamefont {Repp}\ \emph {et~al.}(2005)\citenamefont {Repp},
  \citenamefont {Meyer}, \citenamefont {Stojkovi\ifmmode~\acute{c}\else
  \'{c}\fi{}}, \citenamefont {Gourdon},\ and\ \citenamefont
  {Joachim}}]{Repp2005}%
  \BibitemOpen
  \bibfield  {author} {\bibinfo {author} {\bibfnamefont {Jascha}\ \bibnamefont
  {Repp}}, \bibinfo {author} {\bibfnamefont {Gerhard}\ \bibnamefont {Meyer}},
  \bibinfo {author} {\bibfnamefont {Sladjana~M.}\ \bibnamefont
  {Stojkovi\ifmmode~\acute{c}\else \'{c}\fi{}}}, \bibinfo {author}
  {\bibfnamefont {Andr\'e}\ \bibnamefont {Gourdon}}, \ and\ \bibinfo {author}
  {\bibfnamefont {Christian}\ \bibnamefont {Joachim}},\ }\bibfield  {title}
  {\enquote {\bibinfo {title} {Molecules on insulating films:
  Scanning-tunneling microscopy imaging of individual molecular orbitals},}\
  }\href {\doibase 10.1103/PhysRevLett.94.026803} {\bibfield  {journal}
  {\bibinfo  {journal} {Phys. Rev. Lett.}\ }\textbf {\bibinfo {volume} {94}},\
  \bibinfo {pages} {026803} (\bibinfo {year} {2005})}\BibitemShut {NoStop}%
\bibitem [{\citenamefont {Qiu}\ \emph {et~al.}(2003)\citenamefont {Qiu},
  \citenamefont {Nazin},\ and\ \citenamefont {Ho}}]{Qiu2003}%
  \BibitemOpen
  \bibfield  {author} {\bibinfo {author} {\bibfnamefont {X.~H.}\ \bibnamefont
  {Qiu}}, \bibinfo {author} {\bibfnamefont {G.~V.}\ \bibnamefont {Nazin}}, \
  and\ \bibinfo {author} {\bibfnamefont {W.}~\bibnamefont {Ho}},\ }\bibfield
  {title} {\enquote {\bibinfo {title} {Vibrationally resolved fluorescence
  excited with submolecular precision},}\ }\href {\doibase
  10.1126/science.1078675} {\bibfield  {journal} {\bibinfo  {journal}
  {Science}\ }\textbf {\bibinfo {volume} {299}},\ \bibinfo {pages} {542--546}
  (\bibinfo {year} {2003})}\BibitemShut {NoStop}%
\bibitem [{\citenamefont {Schulz}\ \emph {et~al.}(2013)\citenamefont {Schulz},
  \citenamefont {Drost}, \citenamefont {Hämäläinen},\ and\ \citenamefont
  {Liljeroth}}]{Schulz2013}%
  \BibitemOpen
  \bibfield  {author} {\bibinfo {author} {\bibfnamefont {Fabian}\ \bibnamefont
  {Schulz}}, \bibinfo {author} {\bibfnamefont {Robert}\ \bibnamefont {Drost}},
  \bibinfo {author} {\bibfnamefont {Sampsa~K.}\ \bibnamefont {Hämäläinen}},
  \ and\ \bibinfo {author} {\bibfnamefont {Peter}\ \bibnamefont {Liljeroth}},\
  }\bibfield  {title} {\enquote {\bibinfo {title} {Templated self-assembly and
  local doping of molecules on epitaxial hexagonal boron nitride},}\ }\href
  {\doibase 10.1021/nn404840h} {\bibfield  {journal} {\bibinfo  {journal} {ACS
  Nano}\ }\textbf {\bibinfo {volume} {7}},\ \bibinfo {pages} {11121--11128}
  (\bibinfo {year} {2013})}\BibitemShut {NoStop}%
\bibitem [{\citenamefont {Tautz}(2007)}]{Tautz2007}%
  \BibitemOpen
  \bibfield  {author} {\bibinfo {author} {\bibfnamefont {F.S.}\ \bibnamefont
  {Tautz}},\ }\bibfield  {title} {\enquote {\bibinfo {title} {Structure and
  bonding of large aromatic molecules on noble metal surfaces: The example of
  {PTCDA}},}\ }\href {\doibase https://doi.org/10.1016/j.progsurf.2007.09.001}
  {\bibfield  {journal} {\bibinfo  {journal} {Prog. Surf. Sci.}\ }\textbf
  {\bibinfo {volume} {82}},\ \bibinfo {pages} {479--520} (\bibinfo {year}
  {2007})}\BibitemShut {NoStop}%
\bibitem [{\citenamefont {Lu}\ \emph {et~al.}(2004)\citenamefont {Lu},
  \citenamefont {Grobis}, \citenamefont {Khoo}, \citenamefont {Louie},\ and\
  \citenamefont {Crommie}}]{Lu2004}%
  \BibitemOpen
  \bibfield  {author} {\bibinfo {author} {\bibfnamefont {Xinghua}\ \bibnamefont
  {Lu}}, \bibinfo {author} {\bibfnamefont {M.}~\bibnamefont {Grobis}}, \bibinfo
  {author} {\bibfnamefont {K.~H.}\ \bibnamefont {Khoo}}, \bibinfo {author}
  {\bibfnamefont {Steven~G.}\ \bibnamefont {Louie}}, \ and\ \bibinfo {author}
  {\bibfnamefont {M.~F.}\ \bibnamefont {Crommie}},\ }\bibfield  {title}
  {\enquote {\bibinfo {title} {Charge transfer and screening in individual
  {C}$_{60}$ molecules on metal substrates: A scanning tunneling spectroscopy
  and theoretical study},}\ }\href {\doibase 10.1103/PhysRevB.70.115418}
  {\bibfield  {journal} {\bibinfo  {journal} {Phys. Rev. B}\ }\textbf {\bibinfo
  {volume} {70}},\ \bibinfo {pages} {115418} (\bibinfo {year}
  {2004})}\BibitemShut {NoStop}%
\bibitem [{\citenamefont {Chang}\ \emph {et~al.}(2016)\citenamefont {Chang},
  \citenamefont {Liu}, \citenamefont {Lin}, \citenamefont {Wang}, \citenamefont
  {Zhao}, \citenamefont {Zhang}, \citenamefont {Jin}, \citenamefont {Zhong},
  \citenamefont {Hu}, \citenamefont {Duan}, \citenamefont {Zhang},
  \citenamefont {Fu}, \citenamefont {Xue}, \citenamefont {Chen},\ and\
  \citenamefont {Ji}}]{Chang2016}%
  \BibitemOpen
  \bibfield  {author} {\bibinfo {author} {\bibfnamefont {Kai}\ \bibnamefont
  {Chang}}, \bibinfo {author} {\bibfnamefont {Junwei}\ \bibnamefont {Liu}},
  \bibinfo {author} {\bibfnamefont {Haicheng}\ \bibnamefont {Lin}}, \bibinfo
  {author} {\bibfnamefont {Na}~\bibnamefont {Wang}}, \bibinfo {author}
  {\bibfnamefont {Kun}\ \bibnamefont {Zhao}}, \bibinfo {author} {\bibfnamefont
  {Anmin}\ \bibnamefont {Zhang}}, \bibinfo {author} {\bibfnamefont {Feng}\
  \bibnamefont {Jin}}, \bibinfo {author} {\bibfnamefont {Yong}\ \bibnamefont
  {Zhong}}, \bibinfo {author} {\bibfnamefont {Xiaopeng}\ \bibnamefont {Hu}},
  \bibinfo {author} {\bibfnamefont {Wenhui}\ \bibnamefont {Duan}}, \bibinfo
  {author} {\bibfnamefont {Qingming}\ \bibnamefont {Zhang}}, \bibinfo {author}
  {\bibfnamefont {Liang}\ \bibnamefont {Fu}}, \bibinfo {author} {\bibfnamefont
  {Qi-Kun}\ \bibnamefont {Xue}}, \bibinfo {author} {\bibfnamefont
  {Xi}~\bibnamefont {Chen}}, \ and\ \bibinfo {author} {\bibfnamefont
  {Shuai-Hua}\ \bibnamefont {Ji}},\ }\bibfield  {title} {\enquote {\bibinfo
  {title} {Discovery of robust in-plane ferroelectricity in atomic-thick
  {SnTe}},}\ }\href {\doibase 10.1126/science.aad8609} {\bibfield  {journal}
  {\bibinfo  {journal} {Science}\ }\textbf {\bibinfo {volume} {353}},\ \bibinfo
  {pages} {274--278} (\bibinfo {year} {2016})}\BibitemShut {NoStop}%
\bibitem [{\citenamefont {Ichibha}\ \emph {et~al.}(2017)\citenamefont
  {Ichibha}, \citenamefont {Hou}, \citenamefont {Hongo},\ and\ \citenamefont
  {Maezono}}]{doi:10.1038/s41598-017-01668-6}%
  \BibitemOpen
  \bibfield  {author} {\bibinfo {author} {\bibfnamefont {Tom}\ \bibnamefont
  {Ichibha}}, \bibinfo {author} {\bibfnamefont {Zhufeng}\ \bibnamefont {Hou}},
  \bibinfo {author} {\bibfnamefont {Kenta}\ \bibnamefont {Hongo}}, \ and\
  \bibinfo {author} {\bibfnamefont {Ryo}\ \bibnamefont {Maezono}},\ }\bibfield
  {title} {\enquote {\bibinfo {title} {New insight into the ground state of
  fepc: A diffusion monte carlo study},}\ }\href
  {https://doi.org/10.1038/s41598-017-01668-6} {\bibfield  {journal} {\bibinfo
  {journal} {Sci. Rep.}\ }\textbf {\bibinfo {volume} {7}},\ \bibinfo {pages}
  {2011} (\bibinfo {year} {2017})}\BibitemShut {NoStop}%
\bibitem [{\citenamefont {de~la Torre}\ \emph {et~al.}(2018)\citenamefont
  {de~la Torre}, \citenamefont {Švec}, \citenamefont {Hapala}, \citenamefont
  {Redondo}, \citenamefont {Krejčí}, \citenamefont {Lo}, \citenamefont
  {Manna}, \citenamefont {Sarmah}, \citenamefont {Nachtigallová},
  \citenamefont {Tuček}, \citenamefont {Błoński}, \citenamefont {Otyepka},
  \citenamefont {Zbořil}, \citenamefont {Hobza},\ and\ \citenamefont
  {Jelínek}}]{doi.org/10.1038/s41467-018-05163-y}%
  \BibitemOpen
  \bibfield  {author} {\bibinfo {author} {\bibfnamefont {Bruno}\ \bibnamefont
  {de~la Torre}}, \bibinfo {author} {\bibfnamefont {Martin}\ \bibnamefont
  {Švec}}, \bibinfo {author} {\bibfnamefont {Prokop}\ \bibnamefont {Hapala}},
  \bibinfo {author} {\bibfnamefont {Jesus}\ \bibnamefont {Redondo}}, \bibinfo
  {author} {\bibfnamefont {Ondřej}\ \bibnamefont {Krejčí}}, \bibinfo
  {author} {\bibfnamefont {Rabindranath}\ \bibnamefont {Lo}}, \bibinfo {author}
  {\bibfnamefont {Debashree}\ \bibnamefont {Manna}}, \bibinfo {author}
  {\bibfnamefont {Amrit}\ \bibnamefont {Sarmah}}, \bibinfo {author}
  {\bibfnamefont {Dana}\ \bibnamefont {Nachtigallová}}, \bibinfo {author}
  {\bibfnamefont {Jiří}\ \bibnamefont {Tuček}}, \bibinfo {author}
  {\bibfnamefont {Piotr}\ \bibnamefont {Błoński}}, \bibinfo {author}
  {\bibfnamefont {Michal}\ \bibnamefont {Otyepka}}, \bibinfo {author}
  {\bibfnamefont {Radek}\ \bibnamefont {Zbořil}}, \bibinfo {author}
  {\bibfnamefont {Pavel}\ \bibnamefont {Hobza}}, \ and\ \bibinfo {author}
  {\bibfnamefont {Pavel}\ \bibnamefont {Jelínek}},\ }\bibfield  {title}
  {\enquote {\bibinfo {title} {Non-covalent control of spin-state in
  metal-organic complex by positioning on n-doped graphene},}\ }\href
  {https://doi.org/10.1038/s41467-018-05163-y} {\bibfield  {journal} {\bibinfo
  {journal} {Nat. Commun.}\ }\textbf {\bibinfo {volume} {9}},\ \bibinfo {pages}
  {2831} (\bibinfo {year} {2018})}\BibitemShut {NoStop}%
\bibitem [{\citenamefont {Chang}\ \emph {et~al.}(2020)\citenamefont {Chang},
  \citenamefont {Küster}, \citenamefont {Miller}, \citenamefont {Ji},
  \citenamefont {Zhang}, \citenamefont {Sessi}, \citenamefont {Barraza-Lopez},\
  and\ \citenamefont {Parkin}}]{Chang2020a}%
  \BibitemOpen
  \bibfield  {author} {\bibinfo {author} {\bibfnamefont {Kai}\ \bibnamefont
  {Chang}}, \bibinfo {author} {\bibfnamefont {Felix}\ \bibnamefont {Küster}},
  \bibinfo {author} {\bibfnamefont {Brandon~J.}\ \bibnamefont {Miller}},
  \bibinfo {author} {\bibfnamefont {Jing-Rong}\ \bibnamefont {Ji}}, \bibinfo
  {author} {\bibfnamefont {Jia-Lu}\ \bibnamefont {Zhang}}, \bibinfo {author}
  {\bibfnamefont {Paolo}\ \bibnamefont {Sessi}}, \bibinfo {author}
  {\bibfnamefont {Salvador}\ \bibnamefont {Barraza-Lopez}}, \ and\ \bibinfo
  {author} {\bibfnamefont {Stuart S.~P.}\ \bibnamefont {Parkin}},\ }\bibfield
  {title} {\enquote {\bibinfo {title} {Microscopic manipulation of
  ferroelectric domains in {SnSe} monolayers at room temperature},}\ }\href
  {\doibase 10.1021/acs.nanolett.0c02357} {\bibfield  {journal} {\bibinfo
  {journal} {Nano Lett.}\ }\textbf {\bibinfo {volume} {20}},\ \bibinfo {pages}
  {6590--6597} (\bibinfo {year} {2020})}\BibitemShut {NoStop}%
\bibitem [{\citenamefont {Zhang}\ \emph {et~al.}(2022)\citenamefont {Zhang},
  \citenamefont {Nie}, \citenamefont {Zhang}, \citenamefont {Yuan},
  \citenamefont {Fu},\ and\ \citenamefont {Zhang}}]{Zhang2022}%
  \BibitemOpen
  \bibfield  {author} {\bibinfo {author} {\bibfnamefont {Zhimo}\ \bibnamefont
  {Zhang}}, \bibinfo {author} {\bibfnamefont {Jinhua}\ \bibnamefont {Nie}},
  \bibinfo {author} {\bibfnamefont {Zhihao}\ \bibnamefont {Zhang}}, \bibinfo
  {author} {\bibfnamefont {Yuan}\ \bibnamefont {Yuan}}, \bibinfo {author}
  {\bibfnamefont {Ying-Shuang}\ \bibnamefont {Fu}}, \ and\ \bibinfo {author}
  {\bibfnamefont {Wenhao}\ \bibnamefont {Zhang}},\ }\bibfield  {title}
  {\enquote {\bibinfo {title} {Atomic visualization and switching of
  ferroelectric order in $\beta$-{In}$_2${Se}$_3$ films at the single layer
  limit},}\ }\href {\doibase https://doi.org/10.1002/adma.202106951} {\bibfield
   {journal} {\bibinfo  {journal} {Adv. Mater.}\ }\textbf {\bibinfo {volume}
  {34}},\ \bibinfo {pages} {2106951} (\bibinfo {year} {2022})}\BibitemShut
  {NoStop}%
\bibitem [{\citenamefont {Krems}(2018)}]{2018}%
  \BibitemOpen
  \bibfield  {author} {\bibinfo {author} {\bibfnamefont {Roman~V.}\
  \bibnamefont {Krems}},\ }\enquote {\bibinfo {title} {{DC Stark} effect},}\
  in\ \href {\doibase https://doi.org/10.1002/9781119382638.ch2} {\emph
  {\bibinfo {booktitle} {Molecules in Electromagnetic Fields}}}\ (\bibinfo
  {publisher} {John Wiley \& Sons, Ltd},\ \bibinfo {year} {2018})\
  Chap.~\bibinfo {chapter} {2}, pp.\ \bibinfo {pages} {35--58}\BibitemShut
  {NoStop}%
\bibitem [{\citenamefont {Mugarza}\ \emph {et~al.}(2012)\citenamefont
  {Mugarza}, \citenamefont {Robles}, \citenamefont {Krull}, \citenamefont
  {Koryt\'ar}, \citenamefont {Lorente},\ and\ \citenamefont
  {Gambardella}}]{PhysRevB.85.155437}%
  \BibitemOpen
  \bibfield  {author} {\bibinfo {author} {\bibfnamefont {A.}~\bibnamefont
  {Mugarza}}, \bibinfo {author} {\bibfnamefont {R.}~\bibnamefont {Robles}},
  \bibinfo {author} {\bibfnamefont {C.}~\bibnamefont {Krull}}, \bibinfo
  {author} {\bibfnamefont {R.}~\bibnamefont {Koryt\'ar}}, \bibinfo {author}
  {\bibfnamefont {N.}~\bibnamefont {Lorente}}, \ and\ \bibinfo {author}
  {\bibfnamefont {P.}~\bibnamefont {Gambardella}},\ }\bibfield  {title}
  {\enquote {\bibinfo {title} {Electronic and magnetic properties of
  molecule-metal interfaces: Transition-metal phthalocyanines adsorbed on
  {Ag(100)}},}\ }\href {\doibase 10.1103/PhysRevB.85.155437} {\bibfield
  {journal} {\bibinfo  {journal} {Phys. Rev. B}\ }\textbf {\bibinfo {volume}
  {85}},\ \bibinfo {pages} {155437} (\bibinfo {year} {2012})}\BibitemShut
  {NoStop}%
\bibitem [{\citenamefont {Fern\'andez-Rodr\'{\i}guez}\ \emph
  {et~al.}(2015)\citenamefont {Fern\'andez-Rodr\'{\i}guez}, \citenamefont
  {Toby},\ and\ \citenamefont {van Veenendaal}}]{PhysRevB.91.214427}%
  \BibitemOpen
  \bibfield  {author} {\bibinfo {author} {\bibfnamefont {Javier}\ \bibnamefont
  {Fern\'andez-Rodr\'{\i}guez}}, \bibinfo {author} {\bibfnamefont {Brian}\
  \bibnamefont {Toby}}, \ and\ \bibinfo {author} {\bibfnamefont {Michel}\
  \bibnamefont {van Veenendaal}},\ }\bibfield  {title} {\enquote {\bibinfo
  {title} {Mixed configuration ground state in iron({II}) phthalocyanine},}\
  }\href {\doibase 10.1103/PhysRevB.91.214427} {\bibfield  {journal} {\bibinfo
  {journal} {Phys. Rev. B}\ }\textbf {\bibinfo {volume} {91}},\ \bibinfo
  {pages} {214427} (\bibinfo {year} {2015})}\BibitemShut {NoStop}%
\bibitem [{\citenamefont {Tsukahara}\ \emph {et~al.}(2016)\citenamefont
  {Tsukahara}, \citenamefont {Kawai},\ and\ \citenamefont
  {Takagi}}]{doi:10.1063/1.4940138}%
  \BibitemOpen
  \bibfield  {author} {\bibinfo {author} {\bibfnamefont {Noriyuki}\
  \bibnamefont {Tsukahara}}, \bibinfo {author} {\bibfnamefont {Maki}\
  \bibnamefont {Kawai}}, \ and\ \bibinfo {author} {\bibfnamefont {Noriaki}\
  \bibnamefont {Takagi}},\ }\bibfield  {title} {\enquote {\bibinfo {title}
  {Impact of reduced symmetry on magnetic anisotropy of a single iron
  phthalocyanine molecule on a {Cu} substrate},}\ }\href {\doibase
  10.1063/1.4940138} {\bibfield  {journal} {\bibinfo  {journal} {J. Chem.
  Phys.}\ }\textbf {\bibinfo {volume} {144}},\ \bibinfo {pages} {044701}
  (\bibinfo {year} {2016})}\BibitemShut {NoStop}%
\bibitem [{\citenamefont {Cococcioni}\ and\ \citenamefont
  {de~Gironcoli}(2005)}]{PhysRevB.71.035105}%
  \BibitemOpen
  \bibfield  {author} {\bibinfo {author} {\bibfnamefont {Matteo}\ \bibnamefont
  {Cococcioni}}\ and\ \bibinfo {author} {\bibfnamefont {Stefano}\ \bibnamefont
  {de~Gironcoli}},\ }\bibfield  {title} {\enquote {\bibinfo {title} {Linear
  response approach to the calculation of the effective interaction parameters
  in the $\mathrm{LDA}+\mathrm{U}$ method},}\ }\href {\doibase
  10.1103/PhysRevB.71.035105} {\bibfield  {journal} {\bibinfo  {journal} {Phys.
  Rev. B}\ }\textbf {\bibinfo {volume} {71}},\ \bibinfo {pages} {035105}
  (\bibinfo {year} {2005})}\BibitemShut {NoStop}%
\bibitem [{\citenamefont {Giannozzi}\ \emph {et~al.}(2009)\citenamefont
  {Giannozzi}, \citenamefont {Baroni}, \citenamefont {Bonini}, \citenamefont
  {Calandra}, \citenamefont {Car}, \citenamefont {Cavazzoni}, \citenamefont
  {Ceresoli}, \citenamefont {Chiarotti}, \citenamefont {Cococcioni},
  \citenamefont {Dabo}, \citenamefont {Corso}, \citenamefont {de~Gironcoli},
  \citenamefont {Fabris}, \citenamefont {Fratesi}, \citenamefont {Gebauer},
  \citenamefont {Gerstmann}, \citenamefont {Gougoussis}, \citenamefont
  {Kokalj}, \citenamefont {Lazzeri}, \citenamefont {Martin-Samos},
  \citenamefont {Marzari}, \citenamefont {Mauri}, \citenamefont {Mazzarello},
  \citenamefont {Paolini}, \citenamefont {Pasquarello}, \citenamefont
  {Paulatto}, \citenamefont {Sbraccia}, \citenamefont {Scandolo}, \citenamefont
  {Sclauzero}, \citenamefont {Seitsonen}, \citenamefont {Smogunov},
  \citenamefont {Umari},\ and\ \citenamefont {Wentzcovitch}}]{Giannozzi_2009}%
  \BibitemOpen
  \bibfield  {author} {\bibinfo {author} {\bibfnamefont {Paolo}\ \bibnamefont
  {Giannozzi}}, \bibinfo {author} {\bibfnamefont {Stefano}\ \bibnamefont
  {Baroni}}, \bibinfo {author} {\bibfnamefont {Nicola}\ \bibnamefont {Bonini}},
  \bibinfo {author} {\bibfnamefont {Matteo}\ \bibnamefont {Calandra}}, \bibinfo
  {author} {\bibfnamefont {Roberto}\ \bibnamefont {Car}}, \bibinfo {author}
  {\bibfnamefont {Carlo}\ \bibnamefont {Cavazzoni}}, \bibinfo {author}
  {\bibfnamefont {Davide}\ \bibnamefont {Ceresoli}}, \bibinfo {author}
  {\bibfnamefont {Guido~L}\ \bibnamefont {Chiarotti}}, \bibinfo {author}
  {\bibfnamefont {Matteo}\ \bibnamefont {Cococcioni}}, \bibinfo {author}
  {\bibfnamefont {Ismaila}\ \bibnamefont {Dabo}}, \bibinfo {author}
  {\bibfnamefont {Andrea~Dal}\ \bibnamefont {Corso}}, \bibinfo {author}
  {\bibfnamefont {Stefano}\ \bibnamefont {de~Gironcoli}}, \bibinfo {author}
  {\bibfnamefont {Stefano}\ \bibnamefont {Fabris}}, \bibinfo {author}
  {\bibfnamefont {Guido}\ \bibnamefont {Fratesi}}, \bibinfo {author}
  {\bibfnamefont {Ralph}\ \bibnamefont {Gebauer}}, \bibinfo {author}
  {\bibfnamefont {Uwe}\ \bibnamefont {Gerstmann}}, \bibinfo {author}
  {\bibfnamefont {Christos}\ \bibnamefont {Gougoussis}}, \bibinfo {author}
  {\bibfnamefont {Anton}\ \bibnamefont {Kokalj}}, \bibinfo {author}
  {\bibfnamefont {Michele}\ \bibnamefont {Lazzeri}}, \bibinfo {author}
  {\bibfnamefont {Layla}\ \bibnamefont {Martin-Samos}}, \bibinfo {author}
  {\bibfnamefont {Nicola}\ \bibnamefont {Marzari}}, \bibinfo {author}
  {\bibfnamefont {Francesco}\ \bibnamefont {Mauri}}, \bibinfo {author}
  {\bibfnamefont {Riccardo}\ \bibnamefont {Mazzarello}}, \bibinfo {author}
  {\bibfnamefont {Stefano}\ \bibnamefont {Paolini}}, \bibinfo {author}
  {\bibfnamefont {Alfredo}\ \bibnamefont {Pasquarello}}, \bibinfo {author}
  {\bibfnamefont {Lorenzo}\ \bibnamefont {Paulatto}}, \bibinfo {author}
  {\bibfnamefont {Carlo}\ \bibnamefont {Sbraccia}}, \bibinfo {author}
  {\bibfnamefont {Sandro}\ \bibnamefont {Scandolo}}, \bibinfo {author}
  {\bibfnamefont {Gabriele}\ \bibnamefont {Sclauzero}}, \bibinfo {author}
  {\bibfnamefont {Ari~P}\ \bibnamefont {Seitsonen}}, \bibinfo {author}
  {\bibfnamefont {Alexander}\ \bibnamefont {Smogunov}}, \bibinfo {author}
  {\bibfnamefont {Paolo}\ \bibnamefont {Umari}}, \ and\ \bibinfo {author}
  {\bibfnamefont {Renata~M}\ \bibnamefont {Wentzcovitch}},\ }\bibfield  {title}
  {\enquote {\bibinfo {title} {{QUANTUM} {ESPRESSO}: a modular and open-source
  software project for quantum simulations of materials},}\ }\href {\doibase
  10.1088/0953-8984/21/39/395502} {\bibfield  {journal} {\bibinfo  {journal}
  {J. Phys.: Condens. Matter}\ }\textbf {\bibinfo {volume} {21}},\ \bibinfo
  {pages} {395502} (\bibinfo {year} {2009})}\BibitemShut {NoStop}%
\bibitem [{\citenamefont {Rappe}\ \emph {et~al.}(1990)\citenamefont {Rappe},
  \citenamefont {Rabe}, \citenamefont {Kaxiras},\ and\ \citenamefont
  {Joannopoulos}}]{PhysRevB.41.1227}%
  \BibitemOpen
  \bibfield  {author} {\bibinfo {author} {\bibfnamefont {Andrew~M.}\
  \bibnamefont {Rappe}}, \bibinfo {author} {\bibfnamefont {Karin~M.}\
  \bibnamefont {Rabe}}, \bibinfo {author} {\bibfnamefont {Efthimios}\
  \bibnamefont {Kaxiras}}, \ and\ \bibinfo {author} {\bibfnamefont {J.~D.}\
  \bibnamefont {Joannopoulos}},\ }\bibfield  {title} {\enquote {\bibinfo
  {title} {Optimized pseudopotentials},}\ }\href {\doibase
  10.1103/PhysRevB.41.1227} {\bibfield  {journal} {\bibinfo  {journal} {Phys.
  Rev. B}\ }\textbf {\bibinfo {volume} {41}},\ \bibinfo {pages} {1227--1230}
  (\bibinfo {year} {1990})}\BibitemShut {NoStop}%
\bibitem [{\citenamefont {Perdew}\ \emph {et~al.}(1996)\citenamefont {Perdew},
  \citenamefont {Burke},\ and\ \citenamefont
  {Ernzerhof}}]{PhysRevLett.77.3865}%
  \BibitemOpen
  \bibfield  {author} {\bibinfo {author} {\bibfnamefont {John~P.}\ \bibnamefont
  {Perdew}}, \bibinfo {author} {\bibfnamefont {Kieron}\ \bibnamefont {Burke}},
  \ and\ \bibinfo {author} {\bibfnamefont {Matthias}\ \bibnamefont
  {Ernzerhof}},\ }\bibfield  {title} {\enquote {\bibinfo {title} {Generalized
  gradient approximation made simple},}\ }\href {\doibase
  10.1103/PhysRevLett.77.3865} {\bibfield  {journal} {\bibinfo  {journal}
  {Phys. Rev. Lett.}\ }\textbf {\bibinfo {volume} {77}},\ \bibinfo {pages}
  {3865--3868} (\bibinfo {year} {1996})}\BibitemShut {NoStop}%
\bibitem [{\citenamefont {Grimme}\ \emph {et~al.}(2010)\citenamefont {Grimme},
  \citenamefont {Antony}, \citenamefont {Ehrlich},\ and\ \citenamefont
  {Krieg}}]{doi:10.1063/1.3382344}%
  \BibitemOpen
  \bibfield  {author} {\bibinfo {author} {\bibfnamefont {Stefan}\ \bibnamefont
  {Grimme}}, \bibinfo {author} {\bibfnamefont {Jens}\ \bibnamefont {Antony}},
  \bibinfo {author} {\bibfnamefont {Stephan}\ \bibnamefont {Ehrlich}}, \ and\
  \bibinfo {author} {\bibfnamefont {Helge}\ \bibnamefont {Krieg}},\ }\bibfield
  {title} {\enquote {\bibinfo {title} {A consistent and accurate ab initio
  parametrization of density functional dispersion correction ({DFT-D}) for the
  94 elements {H-Pu}},}\ }\href {\doibase 10.1063/1.3382344} {\bibfield
  {journal} {\bibinfo  {journal} {J. Chem. Phys.}\ }\textbf {\bibinfo {volume}
  {132}},\ \bibinfo {pages} {154104} (\bibinfo {year} {2010})}\BibitemShut
  {NoStop}%
\end{thebibliography}%

\section*{Methods}
An SnTe monolayer was grown by molecular beam epitaxy (MBE) on highly oriented pyrolytic graphite (HOPG) under ultra-high vacuum conditions (UHV, base pressure $\sim1\times10^{-10}$ mbar). HOPG crystal was cleaved and subsequently out-gassed at $\sim300^\circ$C. We deposited SnTe by sublimation from a powder onto the substrate held at $\sim210^\circ$C. The deposition temperature of the SnTe was $\sim550^\circ$C and deposition time was 1 hour. Single, isolated FePc molecules were deposited onto the sample inside the STM at $T=4$ K. FePc monolayer islands were grown by first depositing the FePc molecules onto the substrate at 4 K and then annealing the sample at 200$^\circ$C for 10 minutes.

DFT+U calculations were performed using the Cococcioni and de Gironcoli simplified version \cite{PhysRevB.71.035105} in QUANTUM ESPRESSO package \cite{Giannozzi_2009}, where the Hubbard U parameter for the Fe 3$d$ orbitals was considered to be 4 eV. Electron-ion interactions were represented by ultrasoft pseudopotentials generated with the Rappe-Rabe-Kaxiras-Joannopoulos recipe \cite{PhysRevB.41.1227}. The electronic exchange-correlation potential was calculated using the Perdew-Burke-Ernzerhof (PBE) functional \cite{PhysRevLett.77.3865}, and vdW corrections were taken into account through the empirical DFT-D3 Grimme scheme \cite{doi:10.1063/1.3382344}.  Electronic wave functions were expanded in plane waves with an energy cutoff of 46 Ry, while the cutoff for the charge density was taken to 326 Ry. The atomic positions of both gas phase FePc molecule and FePc + SnTe were optimized until the residual forces were less than 0.001 Ry/a.u. Spin polarization was considered in all calculations where the FePc molecule was present, with a starting magnetization of 2 $\mu_B$ per Fe atom. Different spin configurations were obtained by manipulating the occupation matrix $U$ within the DFT+U method.

\section*{Data availability}
All of the data supporting the findings are available from the corresponding authors upon request.

\section*{Acknowledgements}
This research made use of the Aalto Nanomicroscopy Center (Aalto NMC) facilities and was supported by the European Research Council (ERC-2021-StG no.~101039500 ``Tailoring Quantum Matter on the Flatland'' and ERC-2017-AdG no.~788185 ``Artificial Designer Materials'') and Academy of Finland (Academy professor funding nos.~318995 and 320555, Academy research fellow nos.~331342, 336243 and no.~338478 and 346654). A.S.F. has been supported by the
World Premier International Research Center Initiative (WPI),
MEXT, Japan. We acknowledge the computational resources provided by the Aalto Science-IT project and CSC, Helsinki. 

\section*{Author contributions}
M.A., S.K., and V.V. performed the experiments. M.A. and S.K. analyzed the experimental data. O.J.S. performed the DFT calculations under the supervision of J.L, and A.S.F.
M.A., S.K., and P.L.  designed the experiment. M.A., S.K., O.J.S., P.L., wrote the manuscript, with input from all authors

\section*{Competing interests}
The authors declare no competing interests.

\end{document}